# VR Fire Safety Training Application: A Virtual Reality-Based Training System for Fire Emergency Preparedness

UJWAL M R
Department of MCA
RV College of Engineering
Bengaluru, India
ujwalmr.mca23@rvce.edu.in

Dr. PREETHI N PATIL
Department of MCA
RV College of Engineering
Bengaluru, India
preethinpatil@rvce.edu.in

**Abstract:** Fire safety training is essential for developing emergency response skills and reducing casualties during fire incidents. However, traditional methods such as classroom lectures and fire drills are limited in realism, scalability, and engagement. This paper introduces a Virtual Reality (VR) Fire Safety Training Application designed to address these challenges. Developed using Unity and deployed on Meta Quest 3, the application presents immersive fire scenarios that teach users how to detect hazards, operate fire extinguishers, and evacuate safely. The training application features realistic 3D environments, interactive tasks, performance feedback, and gamified assessments. Results from a user study involving participants show significantly improved knowledge retention, faster evacuation responses, and higher engagement levels compared to traditional methods. The findings suggest that VR-based training can provide an effective and safe alternative to conventional fire safety education.

## I. Introduction

Fire emergencies in buildings, workplaces, and institutions pose significant threats to human life and infrastructure. Despite the implementation of safety regulations and awareness campaigns, many individuals remain unprepared to respond effectively due to limited or ineffective training. Traditional fire safety training methods, such as lectures, posters, and mock drills, often lack realism and fail to simulate the stress and urgency of actual fire scenarios. Furthermore, these methods are difficult to repeat frequently and are constrained by time, location, and logistical limitations.

Virtual Reality has emerged as a promising tool for improving training outcomes through immersive and interactive experiences. By replicating hazardous situations in a safe, controlled environment, VR can enhance user engagement, stimulate real-time decision-making, and enable repeated practice without physical risks. The present research focuses on the development of a VR Fire Safety Training Application aimed at improving fire emergency preparedness through realistic simulation, interactive learning, and performance evaluation. The application is implemented using the Unity game engine and deployed on the Meta Quest 3 VR headset, a standalone device known for its portability and accessibility.

The objectives of this work include the design of immersive fire safety scenarios, the integration of user feedback and assessment modules, and the evaluation of training effectiveness compared to traditional methods. This paper is structured as follows: Section I is the introduction of what the project is about. Section II presents related work and literature survey done. Section III discusses the methodology, Section IV describes results and discussion, Section V concludes the paper and outlines future enhancements. Section VI consists of all the references.

## II. Literature Review

Virtual Reality (VR) has become an increasingly valuable tool for hands-on training in areas like aviation, healthcare, and industrial safety. Its ability to simulate real-world situations in a safe, controlled space allows learners to practice tasks, make decisions, and build confidence—all without facing real danger. Researchers like Witmer and Singer emphasized the role of immersion and presence in VR, showing how these elements boost focus and learning effectiveness.

In safety-related fields, particularly construction and industrial environments, VR has helped improve how trainees identify hazards and respond to emergencies. Repetitive learning in realistic scenarios has been shown to lead to better long-term retention and decision-making. When it comes to fire safety, most VR solutions so far have concentrated on evacuation drills. These experiences guide users along escape routes and build basic emergency awareness. For example, Park et al. developed a fire evacuation simulator that improved user's navigation and response times. Chen et al. took things further by including decision-based scenarios where user's actions affect outcomes.

Despite these advances, many existing systems have clear limitations. They often lack realistic interactivity like physically handling a fire extinguisher or facing branching scenarios that change based on the user's choices. On top of that, most rely on expensive PC-based VR setups, which makes them harder to deploy in schools or small organizations.

Our work builds on these foundations by offering a more interactive and accessible approach. The VR Fire Safety Training Application is designed specifically for the standalone Meta Quest 3 headset, eliminating the need for a tethered setup. It includes fire suppression mechanics, customizable training scenarios, and performance tracking—making fire safety training more engaging, practical, and adaptable for different learning environments.

### III. Methodology

The architecture of the VR Fire Safety Training Application is structured around three key components: scene simulation, user interaction with fire scenarios, and performance assessment.

**1.** The scene simulation is responsible for creating immersive virtual environments that replicate real-world fire-prone settings. The application features a distinct laboratory environment, presenting different fire hazards such as electrical fires, normal fires, and chemical spills. These environments are designed using Unity's 3D engine, integrating lighting, textures to enhance realism. Users are placed directly into these environments with minimal instructions, encouraging instinctive responses and critical thinking.

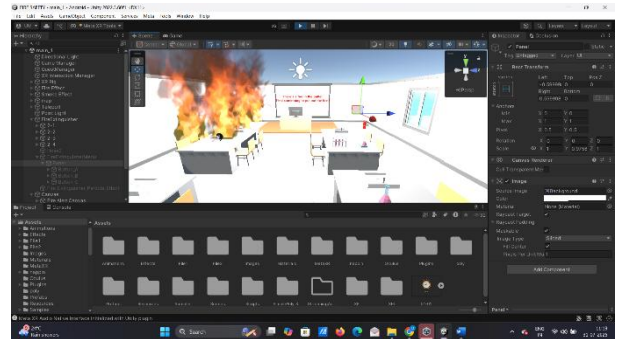

Fig 3.1.  Scene building

Fire behaviour is driven by time-based logic to ensure controlled and predictable training scenarios. Rather than complex physics models. The scene starts with a small fire, and additional fires are programmed to activate at predefined time intervals—simulating the natural escalation of a fire incident.

Fire starts at time 0 seconds

• At time 10 seconds, fire spreads to a new object

• At time 20 seconds, fire spreads further

• At time 30 seconds, fire spreads again

• At time 40 seconds, fire reaches maximum spread

Fire suppression is also modelled with a time-based decay

**FireIntensity(t) = MaxIntensity - (ExtinguishRate × t)**

Where:

- FireIntensity(t) reduces over time as the extinguisher is applied
- ExtinguishRate is a fixed value based on extinguisher type
- Fire is extinguished when FireIntensity(t) ≤ 0

**2.** The user interaction enables the trainee to engage with the environment using VR controllers. Participants must assess the situation, identify the type and source of the fire, locate and operate the appropriate fire extinguisher and navigate safe evacuation routes. In some scenarios, users may also be required to assist virtual bystanders. Fire behaviour is simulated through particle systems, animation controllers and smoke, and alarm sounds. Interactive elements such as fire extinguishers, doors are scripted to respond accurately to user input.

Users interact using VR controllers to extinguish fire. To simulate effectiveness, three main simplified physics rules are applied

**Spray Effectiveness (E)**

$$E = Cos(\theta) * (1 - d/d_{max})$$

- $\theta$: angle between extinguisher spray and fire source
- d: distance between user and fire
- $d_{max}$: maximum effective range (e.g., 3 meters)
- E: Effectiveness

**Extinguishing Progress (P)**

$$P = E * t_{spray}$$

- $t_{spray}$: time user holds the trigger while aiming correctly
- Higher P values reduce the fire size

**Extinguishing Threshold**: Fire is considered extinguished if cumulative P is greater than or equal to fire intensity threshold

$$P \geq \text{fire intensity threshold}$$

These formulas help simulate a realistic extinguishing process where aim, distance, and duration matter.

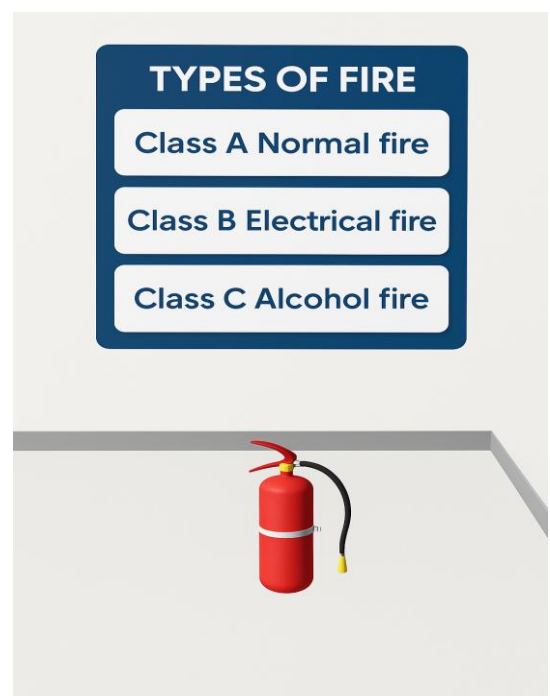

Fig 3.2. Menu for selecting Extinguisher type

**3.** After each training session, the performance assessment module evaluates the user based on several criteria: reaction time, correct use of firefighting equipment, hazard recognition, and adherence to proper evacuation protocols. Results are compiled into a visual feedback report, offering a breakdown of scores across key performance areas including hazard identification, extinguisher handling, and evacuation strategy.

A performance evaluation report is shown, analysing the following parameters

- Response Time
- Accuracy of Aiming
- Correct Usage of Fire Extinguisher
- Completion of Evacuation Steps

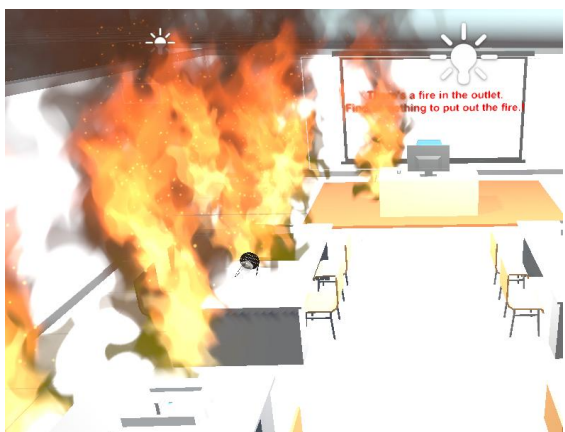

Figure 3.3. Fire simulation

Scores are calculated using logged values, for example:

$$\text{Aiming Score} = \sum F/n \times 100$$

Where,

- F = Number of spray hits directly focused at the fire base (effective sprays)
- N = Total number of spray samples (attempts) recorded during the simulation

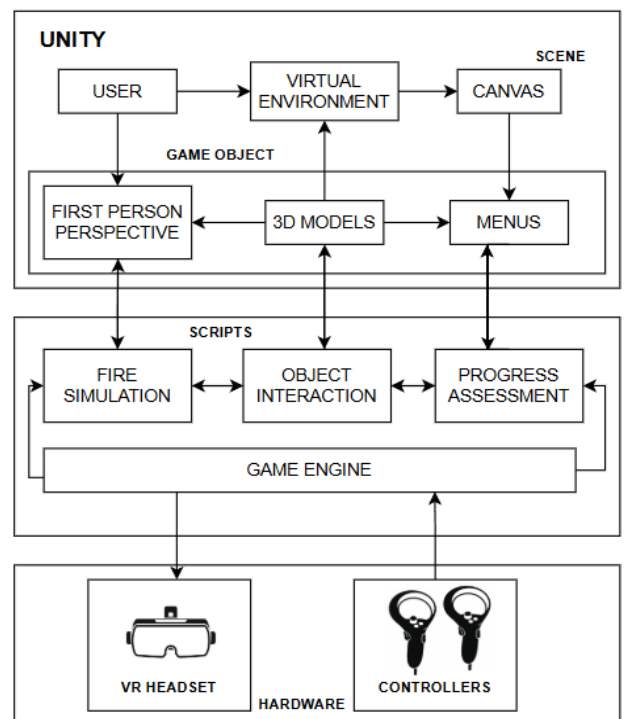

Figure 3.4. Architecture Diagram

## IV. Results and Discussion

This section explores how users progressed and improved while using the VR Fire Safety Training Application. A total of 5 individuals participated in the evaluation, each completing six simulation attempts.

The goal was to assess how their response time, hazard recognition, and extinguisher usage evolved with practice.

We focused on tracking the time taken by the users

- Time Taken (in seconds): This represents the total time a user needed to extinguish the fire and exit the environment.

To better understand skill progression, the attempts were divided into three learning stages

- Initial Phase (Attempts 1 & 2): Gauged the participant's natural responses with minimal prior exposure to the VR environment.
- Improvement Phase (Attempts 3 & 4): Showed the learning curve as users became more familiar and began to respond more effectively.
- Advanced Phase (Attempts 5 & 6): Reflected refined performance, indicating improved decision-making, confidence, and speed.

**Initial: Attempts 1 and 2**

| User | Attempt1 (Time) | Attempt2 (Time) |
|---|---|---|
| U1 | 46s | 41s |
| U2 | DNF | 52s |
| U3 | 48s | DNF |
| U4 | 39s | 36s |
| U5 | DNF | 44s |
| U6 | 34s | 33s |
| U7 | 57s | 55s |
| U8 | DNF | DNF |
| U9 | 41s | 41s |
| U10 | 36s | 35s |

Table 4.1. User data on attempts 1 & 2

**Observation**

In the initial phase of the task, several users faced difficulties, resulting in a total of six incomplete attempts across the first two rounds. One participant, U8, was unable to finish both attempts, while U2, U3, and U5 each had a single incomplete run. On the other hand, users like U4, U6, and U10 showed a strong grasp of the process early on, managing the task with confidence and control. For those who successfully completed the attempts, the average completion time was approximately 43.9 seconds.

**Improvement: Attempts 3 and 4**

| User | Attempt3 (Time) | Attempt4 (Time) |
|---|---|---|
| U1 | 37s | 32s |
| U2 | 45s | DNF |
| U3 | 42s | 35s |
| U4 | 30s | 26s |
| U5 | 39s | 34s |
| U6 | 30s | 28s |
| U7 | DNF | 45s |
| U8 | 47s | DNF |
| U9 | 41s | 40s |
| U10 | 32s | 30s |

Table 4.2. User data on attempts 3 & 4

**Observation**

Most participants showed noticeable improvement during this phase, although a few still encountered issues, leading to three incomplete attempts, specifically by U2 in Attempt 4, U7 in Attempt 3, and U8 again in Attempt 4. Despite earlier setbacks, users such as U3, U5, and U8 made commendable progress and managed to perform better. Meanwhile, U4, U6, and U10 maintained steady and reliable performance throughout. Among the users who completed the tasks, the average completion time was around 36.5 seconds.

**Advanced: Attempts 5 and 6**

| User | Attempt5 (Time) | Attempt6 (Time) |
|---|---|---|
| U1 | 28s | 23s |
| U2 | 35s | 31s |
| U3 | 33s | 29s |
| U4 | 24s | 20s |
| U5 | 31s | 27s |
| U6 | 29s | 28s |
| U7 | 42s | 40s |
| U8 | 30s | 26s |
| U9 | 40s | 30s |
| U10 | 29s | 27s |

Table 4.3. User data on attempts 5 & 6

**Observation**

In the advanced stage, all users were able to complete both attempts. Overall performance improved significantly, with users like U4, U1, and U10 consistently demonstrating strong control and efficiency. U2 and U3 bounced back well from earlier struggles, showing clear progress, while U7 remained slightly behind the rest in terms of speed and execution. For those who completed the tasks, the average time recorded was approximately 29.2 seconds.

Overall, the collected data shows noticeable improvement in efficiency (time taken). These findings support the idea that VR-based training offers an engaging and effective method for building real-world emergency response skills in a safe and repeatable environment.

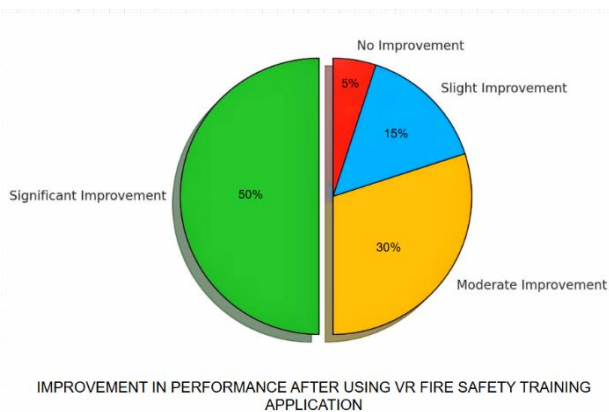

Fig 4.2. Visualisation of performance

## V. Conclusion

This research presents a Virtual Reality-based Fire Safety Training Application designed to deliver immersive, interactive, and effective emergency preparedness training. The system allows users to practice fire response techniques in a safe yet realistic environment, enhancing their understanding and reaction skills. Evaluation results show that VR training leads to better knowledge retention, faster response times, and higher user satisfaction compared to traditional methods.

Future development will focus on expanding the variety of training environments, integrating multiplayer functionality for group evacuations, incorporating voice command recognition, and adding adaptive learning features powered by AI. Further studies involving larger and more diverse user groups will be conducted to validate the generalizability of the results. With continued improvements, the VR Fire Safety Training Application has the potential to become a standard tool in educational and organizational safety programs.